\documentclass[11pt]{article}
\usepackage[pctex32]{graphics}

\def\ben{\begin{equation}}
\def\een{\end{equation}}

\def\nn{\nonumber} \def\bd{\begin{document}} \def\ed{\end{document}}
\def\ds{\documentstyle} \let\fr=\frac \let\bl=\bigl \let\br=\bigr
\let\Br=\Bigr \let\Bl=\Bigl
\let\bm=\bibitem
\let\na=\nabla
\let\pa=\partial \let\ov=\overline
\newcommand{\be}{\begin{equation}}
\newcommand{\ee}{\end{equation}}
\def\ba{\begin{array}}
\def\ea{\end{array}}
\def\ft#1#2{{\textstyle{\frac{\scriptstyle #1}{\scriptstyle #2} } }}
\def\fft#1#2{{\frac{#1}{#2}}}
\def\del{\partial}
\def\vp{\varphi}
\def\sst#1{{\scriptscriptstyle #1}}
\def\oneone{\rlap 1\mkern4mu{\rm l}}
\def\td{\tilde}
\def\wtd{\widetilde}
\def\ie{{\it i.e.\ }}
\def\dalemb#1#2{{\vbox{\hrule height .#2pt
        \hbox{\vrule width.#2pt height#1pt \kern#1pt
                \vrule width.#2pt}
        \hrule height.#2pt}}}
\def\square{\mathord{\dalemb{6.8}{7}\hbox{\hskip1pt}}}
\newcommand{\ho}[1]{$\, ^{#1}$}
\newcommand{\hoch}[1]{$\, ^{#1}$}
\newcommand{\bea}{\setlength\arraycolsep{2pt} \begin{eqnarray}}
\newcommand{\eea}{\end{eqnarray}}
\newcommand{\ra}{\rightarrow}
\newcommand{\lra}{\longrightarrow}
\newcommand{\Lra}{\Leftrightarrow}
\newcommand{\bp}{\tilde \beta^\prime}
\newcommand{\tr}{{\rm tr} }
\newcommand{\Tr}{{\rm Tr} }
\def\0{{\sst{(0)}}}
\def\1{{\sst{(1)}}}
\def\2{{\sst{(2)}}}
\def\3{{\sst{(3)}}}
\def\4{{\sst{(4)}}}
\def\5{{\sst{(5)}}}
\def\6{{\sst{(6)}}}
\def\7{{\sst{(7)}}}
\def\8{{\sst{(8)}}}
\def\m{{\sst{(m)}}}
\def\n{{\sst{(n)}}}
\def\cA{{{\cal A}}}
\def\cB{{{\cal B}}}
\def\cF{{{\cal F}}}
\def\cG{{{\cal G}}}
\def\cH{{{\cal H}}}
\def\tV{\widetilde V}
\def\tW{\widetilde W}
\def\tH{\widetilde H}
\def\tE{\widetilde E}
\def\tF{\widetilde F}
\def\tA{\widetilde A}
\def\im{{{\rm i}}}
\def\tY{{{\wtd Y}}}
\def\ep{{\epsilon}}
\def\vep{{\varepsilon}}
\def\bD{{{\bar D}}}
\def\R{{{\mathbb R}}}
\def\C{{{\mathbb C}}}
\def\H{{{\mathbb H}}}
\def\CP{{{\mathbb C}{\mathbb P}}}
\def\RP{{{\mathbb R}{\mathbb P}}}
\def\Z{{{\mathbb Z}}}
\def\bA{{{\mathbb A}}}
\def\bB{{{\mathbb B}}}
\def\bC{{{\mathbb C}}}
\def\bD{{{\mathbb D}}}
\def\bE{{{\mathbb E}}}
\def\bZ{{{\mathbb Z}}}
\def\Re{{{\frak{Re}}}}
\def\Im{{{\frak{Im}}}}
\def\cosec{{\,\hbox{cosec}\,}}
\def\Gm{{\Gamma_{\!\! -}}}
\def\Gp{{\Gamma_{\!\! +}}}
\def\stan{{standard }}
\def\nonstan{{supernumerary }}
\def\p{{\partial}}
\def\kdel#1{{\fft{\del}{\del#1}}}

\def\bog{{Bogomolny }}
\def\om{{\omega}}

\newcommand{\nnr}{\nonumber \\}
\newcommand{\pd}{\partial}
\newcommand{\ud}{\textrm{d}}
\newcommand{\dTH}{T^{\prime \, 0}_\textrm{H}}
\newcommand{\dOi}{\Omega^{\prime \, 0}_i}
\newcommand{\bx}{{\bf x}}

\begin{document}

\vspace{5mm}
\begin{center}
{\Large \bf ADM  mass and  quasilocal energy   of black hole in
the deformed Ho\v{r}ava-Lifshitz gravity } \vspace{12mm}

{\large   Yun Soo Myung \footnote{e-mail
 address: ysmyung@inje.ac.kr}}
 \\
\vspace{10mm} {\em Institute of Basic Science and School of
Computer Aided Science \\ Inje University, Gimhae 621-749, Korea}
\end{center}

\begin{center}

\underline{Abstract}
\end{center}

Inspired by the Einstein-Born-Infeld black hole, we introduce the
isolated horizon to study the  Kehagias-Sfetsos (KS) black hole in
the deformed Ho\v{r}ava-Lifshitz  gravity. This is because the KS
black hole is more close to the Einstein-Born-Infeld black hole
than the Reissner-Nordstr\"om black hole.
  We find the  horizon and ADM masses  by using
  the first law of thermodynamics and the area-law entropy.
 The mass parameter $m$  is identified with  the
quasilocal energy at infinity. Accordingly, we discuss the phase
transition between the KS and Schwarzschild black holes by
considering the heat capacity and free energy.

\vspace{15pt}

\thispagestyle{empty}





\newpage
\section{Introduction}
Recently Ho\v{r}ava has proposed a renormalizable theory of
gravity at a Lifshitz point~\cite{ho1,ho2},  which  may be
regarded as a UV complete candidate for general relativity. At
short distances the theory of  Ho\v{r}ava-Lifshitz (HL) gravity
describes interacting non-relativistic gravitons and is supposed
to be power counting renormalizable in (1+3) dimensions. Recently,
its black hole solutions have  been intensively investigated
~\cite{LMP,CCO1,CLS,CY,MK,CCO2,KS,Myung,park,Gho09,CL,PW,lkme}.

Concerning the static spherically symmetric (SSS) solutions,
L\"u-Mei-Pope (LMP) have obtained the black hole solution with
dynamical parameter $\lambda$~\cite{LMP} and topological black
holes were found in \cite{CCO1}. Its thermodynamics were studied
in \cite{MK,CCO2} but there remain unclear issues in defining the
ADM mass and entropy because its asymptotes are Lifshitz.

On the other hand, Kehagias and Sfetsos (KS)  have found the
``$\lambda=1$" black hole solution in asymptotically flat
spacetimes using the deformed HL gravity with parameter
$\omega$~\cite{KS}. Its thermodynamics seemed to be nicely defined
when using the first law of thermodynamics in Ref.\cite{Myung}.
However, the entropy takes a very unusual form as $S=A/4+
(\pi/\omega)\ln[A/4]$\cite{CL}. Thus, one has to explain why a
logarithmic term $(\pi/\omega)\ln[A/4]$ appears for the entropy of
black hole in the deformed HL gravity\cite{myungent,CO}. This term
arises because one has used the first law of $dS=dm/T_H$  to
derive the entropy, provided that the Hawking temperature  $T_H$
and the mass $m$ are known. Indeed, the mass $m$ was defined
naively by the condition of the zero metric function $f_{KS}=0$.
Actually, $m$ is not the Arnowitt-Deser-Misner (ADM) mass
$M_{ADM}$ defined at infinity because the metric function $f_{KS}$
is  different from the Reissner-Norstr\"om (RN) black hole, but it
is similar to $f_{BI}$ of the Einstein-Born-Infeld (EBI) black
hole. Here we will identify the mass parameter $m$ with  the
quasilocal energy $E(\infty)$ at infinity. However, for the
Schwarzschild and RN black holes, their ADM masses are just
quasilocal energies at infinity.

Introducing the isolated horizon formalism~\cite{ACS}, one may
resolve the unsatisfactory and uncomplete description of the KS
black hole given by concepts such as ADM mass and event horizon.
This formalism provides a more complete description of what
happens in the neighborhood of the horizon. In this formalism, one
considers spacetimes with an interior boundary, which satisfy
quasilocal boundary condition, insuring that the horizon remains
isolated. The boundary condition means that quasilocal charges
could be defined at horizon, which remain constant in time. These
charges include  horizon mass, horizon electric charge, and
horizon magnetic charge. Importantly, the first law of black hole
thermodynamics for quantities defined only at horizon arises
natually, as part of the requirements of a consistent Hamiltonian
formulation.   In addition, Ashteker-Corichi-Sudarsky (ACS)
conjecture on the relation between the colored black holes and
their solitonic analogs implies that the ADM mass consists of two
contributions: black hole horizon and solitonic residue. Hence,
the colored black holes with index $n$ (EBI black hole with index
$b^2$~\cite{Breton}) can be regarded as bound states of ordinary
black holes and their solitons. We insist  that the isolated
horizon formalism is also applicable to the KS black hole.

Comparing the KS black hole $(m,\omega)$ with the EBI black hole
$(M,Q,b)$, one observes an apparent correspondence such that
$m\leftrightarrow Q^2$ (magnetic charge) and $\omega
\leftrightarrow b^2$ (non-linear coupling parameter). This implies
that the EBI black hole may play a role in understanding the KS
black hole from the deformed HL gravity. At infinity, the EBI
black hole $(M,Q,b^2)$ is indistinguishable from the RN black hole
$(M,Q)$, which implies that $b^2$ is considered as a free
parameter like  color index $n$. Similarly, at infinity, the KS
black hole $(m,\omega)$ is indistinguishable form the
Schwarzschild black hole $(m)$. Hence, we have an index relation
of $n\sim b^2\sim \omega$.

Furthermore, it was well known that many different kinds of black
holes from string theories have the Bekenstein-Hawking entropy of
$S_{BH}=A/4$~\cite{Ahar}. Hence, it would be better  to use the
Bekenstein-Hawking entropy to derive the horizon mass and ADM mass
via the first law of  $dM_{h}=T_H dS_{BH}$.

In this work, we obtain  the horizon and ADM mass of KS  black
hole in the deformed HL gravity. In deriving these masses, we use
the first law of thermodynamics and  the Bekenstein-Hawking
entropy. Also, we confirmed that the horizon  mass satisfies the
modified Smarr formula.

 \section{HL gravity}
Introducing the ADM formalism where the metric is parameterized
\be ds_{ADM}^2= - N^2  dt^2 + g_{ij} \Big(dx^i - N^i dt\Big)
\Big(dx^j - N^j dt\Big)\,, \ee
the Einstein-Hilbert action can be expressed as
\be \label{Eins} S_{EH} = \fft{1}{16\pi G} \int d^4x \sqrt{g} N
\Big[K_{ij} K^{ij} - K^2 + R - 2\Lambda\Big], \ee
where $G$ is Newton's constant and extrinsic curvature $K_{ij}$
takes the form
\be K_{ij} = \fft{1}{2N} \Big(\dot g_{ij} - \nabla_i N_j -
\nabla_j N_i\Big)\,. \ee
Here, a dot denotes a derivative with respect to $t$. An action of
the non-relativistic renormalizable gravitational theory  is given
by~\cite{ho1} \be S_{HL}=\int dtd^3x \Big[{\cal L}_K + {\cal
L}_V\Big],  \label{action1} \ee where the kinetic Lagrangian  is
given by \be {\cal L}_K =\frac{2}{\kappa^2}\sqrt{g} N K_{ij}{\cal
G}^{ijkl}K_{kl}= \frac{2}{\kappa^2}\sqrt{g}
N\Big(K_{ij}K^{ij}-\lambda K^2\Big), \ee with the DeWitt metric
 \be {\cal G}^{ijkl}=
\frac{1}{2}\Big(g^{ik}g^{jl}-g^{il}g^{jk}\Big)-\lambda
g^{ij}g^{kl} \ee
 and its inverse metric
 \be {\cal
G}_{ijkl}=\frac{1}{2}\Big(g_{ik}g_{jl}-g_{il}g_{jk}\Big)-\frac{\lambda}{3\lambda-1}g_{ij}g_{kl}.\ee
The potential Lagrangian is determined by the detailed balance
condition as \bea {\cal L}_V=-\frac{\kappa^2}{2}\sqrt{g}N
E^{ij}{\cal G}_{ijkl}E^{kl}&=&
\sqrt{g}N\Bigg\{\frac{\kappa^2\mu^2}{8(1-3\lambda)}\Big(\frac{1-4\lambda}{4}R^2
+\Lambda_WR-3\Lambda_W^2\Big)\nn \\
 &-&\frac{\kappa^2}{2w^4} \left(C_{ij}
-\frac{\mu w^2}{2}R_{ij}\right) \left(C^{ij} -\frac{\mu
w^2}{2}R^{ij}\right) \Bigg\}.\label{action2} \eea Here the $E$
tensor is defined by \be E^{ij}=\frac{1}{w^2} C^{ij}-\frac{\mu}{2}
\Big(R^{ij}-\frac{R}{2} g^{ij}+\Lambda_Wg^{ij}\Big) \ee with the
Cotton tensor $C_{ij}$ \be
C^{ij}=\frac{\epsilon^{ik\ell}}{\sqrt{g}}\nabla_k\left(R^j{}_\ell
-\frac14R\delta_\ell^j\right).\label{def.K.C} \ee  Explicitly,
$E_{ij}$ could be derived  from the Euclidean topologically
massive gravity \be E^{ij}=\frac{1}{\sqrt{g}} \frac{\delta
W_{TMG}}{\delta g_{ij}} \ee with \be W_{TMG}=\frac{1}{w^2} \int
d^3 x \epsilon^{ikl}\Big(\Gamma^m_{il}\partial_j
\Gamma^l_{km}+\frac{2}{3} \Gamma^n_{il} \Gamma^l_{jm}
\Gamma^m_{kn} \Big)- \mu \int d^3x \sqrt{g}(R-2\Lambda_W), \ee
where $\epsilon^{ikl}$ is a tensor density with
$\epsilon^{123}=1$.

In the IR limit,  comparing ${\cal L}_0$ with Eq.(\ref{Eins}) of
general relativity, the speed of light, Newton's constant and the
cosmological constant are given by
\be c=\fft{\kappa^2\mu}{4}
\sqrt{\fft{\Lambda_W}{1-3\lambda}}\,,\qquad
G=\fft{\kappa^2}{32\pi\,c}\,,\qquad \Lambda_{\rm cc}=\ft32
\Lambda_W\,.\label{cg} \ee The equations of motion were derived in
\cite{cos1} and \cite{LMP}. We would like to mention that the IR
vacuum of this theory is anti-de Sitter (AdS$_4$) spacetimes.
Hence, it is interesting to take a limit of the theory, which may
lead to  a Minkowski vacuum in the IR sector. To this end, one may
deform the theory by introducing a soft violation term of
``$\mu^4R$" $(\tilde{{\cal L}}_V={\cal L}_V+\sqrt{g}N \mu^4R)$ and
then, take the $\Lambda_W \to 0$ limit~\cite{KS}. We call this as
the ``deformed HL gravity". This theory  does not alter the UV
property of the HL gravity, while it changes the IR property. That
is, there exists a Minkowski vacuum, instead of an AdS vacuum. In
the IR limit, the speed of light and Newton's constant are given
by
\be c^2=\fft{\kappa^2\mu^4}{2},~ G=\fft{\kappa^2}{32\pi\,c},
~\lambda=1.\label{kh} \ee

\section{ KS black hole and old thermodynamics}
A static spherically symmetric (SSS) solution to the deformed HL
gravity was obtained by considering the line element
\begin{equation}
ds_{SSS}^2 = -N(r)^2 dt^2 + \frac{dr^2}{f(r)} + r^2 \left ( d
\theta^2 + \sin^2 \theta d \phi^2 \right ) . \label{sss}
\end{equation}
For this purpose, we choose the case of  $K_{ij}=0$ and
$C_{ij}=0$. Actually, the above SSS solution could be  derived
from the deformed potential Lagrangian given by \be \label{depot}
\tilde{{\cal L}}_{V}=\mu^4\sqrt{g}N\Bigg[R
+\frac{1}{2\omega}\frac{4\lambda-1}{3\lambda-1}R^2-\frac{2}{\omega}R_{ij}R_{ij}\Bigg],
\ee where an important  parameter, \be
\omega=\frac{16\mu^2}{\kappa^2} \ee specifies the deformed HL
gravity. Hence, it is emphasized that we have relaxed both the
projectability restriction and detailed balance
condition~\cite{ho1,muko}, since
 the lapse function $N(r)$  depends on the spatial
coordinate $r$ as well as a soft violation term of $\mu^4 R$ is
included. Substituting the metric ansatz (\ref{sss}) into
$\tilde{{\cal L}}_V$, one has the reduced Lagrangian \be
\label{react} \tilde{{\cal L}}_V=\frac{ \mu^4 N }{\sqrt{f}}\Bigg[
\frac{\lambda-1}{2\omega} f'^2 - \frac{2\lambda (f-1)}{\omega
r}f'+ \frac{(2\lambda-1)(f-1)^2}{ \omega r^2}-2(1 - f - r
f')\Bigg]. \ee

For $\lambda=1$,  the KS  solution is given by~\cite{KS}
\begin{equation}
f_{KS} =N^2_{KS}= 1 + \omega r^2 \left ( 1 - \sqrt{1 + \frac{4
m}{\omega r^3}}\right ) \label{KSs}
\end{equation}
where $m$ is a mass parameter. In the limit of $\omega \to \infty$
($\kappa^2 \to 0$ and thus, $\mu^4R$ dominates), it reduces to the
Schwarzschild metric function \be \label{sch}
f_{s}(r)=1-\frac{2m}{r}. \ee
  From the condition of $f_{KS}(r_\pm)=0$, the
outer (inner) horizons are given by \be \label{sol} r_\pm=m\pm
\sqrt{m^2-\frac{1}{2\omega}} \ee which takes the same form as in
the RN hole obtained from Einstein-Maxwell action (linear
electrodynamics) \be r^{RN}_\pm=M\pm \sqrt{M^2-Q^2}\ee when
considering a naive correspondence of \be \label{naivc} m
\leftrightarrow M,~~ \frac{1}{2\omega} \leftrightarrow Q^2. \ee
 In order
to have a black hole solution, it requires that the mass parameter
satisfies the following bound, \be m^2 \ge \frac{1}{2\omega}. \ee

Based on the assumption that the mass parameter $m$ from the
condition of $f_{KS}=0$ could represent the ADM mass,
thermodynamic quantities of Hawking temperature and heat capacity
for the KS black hole were derived as~\cite{Myung} \be
\label{oldt} m(r_\pm)=\frac{1+2\omega r_\pm^2}{4\omega
r_\pm},~~T_H=\frac{2\omega r_+^2-1}{8\pi r_+(\omega
r_+^2+1)},~~C_\omega=-\frac{2\pi}{\omega}\Bigg[\frac{(\omega
r_+^2+1)^2(2\omega r_+^2-1)}{2\omega^2 r_+^4-5\omega
r_+^2-1}\Bigg]. \ee Using the first law of thermodynamics, the
entropy was calculated as \be S=\int dr_+
\Bigg[\frac{1}{T_H}\frac{\partial m}{\partial r_+}\Bigg]+S_0,\ee
which leads to~\cite{CL} \be \label{entropy}
S=\frac{A}{4}+\frac{\pi}{\omega}\ln\Big[\frac{A}{4}\Big]\ee with
$A/4=\pi r_+^2 $ and $S_0=\pi \ln \pi/\omega$. We note that in the
limit of $\omega \to \infty$, Eq. (\ref{entropy}) reduces to the
Bekenstein-Hawking entropy of Schwarzschild black hole as \be
S_{BH}=\frac{A}{4}. \ee If the entropy (\ref{entropy}) is correct,
the logarithmic term should represent a feature of KS black hole
in the deformed HL gravity.  However, there was  no way to explain
the appearance of logarithmic term unless either quantum
correction or thermodynamic correction is
considered~\cite{myungent}.

\section{Einstein-Born-Infeld black holes}
First of all, we expand the metric function for large $r$ as \bea
\label{kss}f_{KS}&\simeq&1+\Bigg(-\frac{2m}{r}+\frac{2m^2}{\omega
r^4}-\frac{4m^3}{\omega^2r^7}+\frac{10m^4}{\omega^3r^{10}}-\frac{28m^5}{\omega^4
r^{13}}+\cdots\Bigg)
\\
&\equiv& 1-\frac{2\tilde{m}(r)}{r},
 \eea
where the mass function $\tilde{m}(r)$ is introduced to take into
account the whole $m$-dependent terms. In the limit of $\omega \to
\infty$, it is obvious that $f_{KS} \to f_s$. We note that the
absence of $1/r^2$-term implies that the deformed HL gravity is a
purely gravity theory. Also, Eq.(\ref{kss})  shows clearly a
different behavior from the RN metric function \be \label{RN}
f_{RN}=1-\frac{2M}{r}+\frac{Q^2}{r^2}. \ee Hence, the naive
correspondence (\ref{naivc}) is questionable.   In order to find a
proper case, one introduces a (3+1)-gravity coupled with nonlinear
electrodynamics known as the Einstein-Born-Infeld (EBI)
action~\cite{Breton}
\begin{equation}\label{action}
S=\int d^4x \sqrt{-g}\Bigg[\frac{R_4}{16\pi
G}+L(P,\tilde{Q})\Bigg],
\end{equation}
where the Born-Infeld Lagrangian is given by
\begin{equation}
L(P,\tilde{Q})=-\frac{P^{\mu\nu}F_{\mu\nu}}{2}+K(P,\tilde{Q}),
\end{equation}
with the structural function $K(P,\tilde{Q})$ \be K(P,\tilde{Q})=
b^2\Bigg(1-\sqrt{1-\frac{2P}{b^2}+\frac{\tilde{Q}^2}{b^4}}\Bigg)
\ee with $P$ and $\tilde{Q}$ the invariants of $P_{\mu\nu}$.
 Here, the constant $b^2$ is the Born-Infeld parameter.
 We introduce the line
element with the metric function $f_{BI}(r)$ as follows:
\begin{equation} \label{metric}
ds_{BI}^2 = - f_{BI}(r) dt^2 + f_{BI}(r)^{-1} dr^2 + r^2 \Big(
d\theta^2+\sin^2\theta d\phi^2\Big).
\end{equation}
Choosing  the SSS background (\ref{metric}), the electrically
(magnetically) charged solution is obtained by  taking
\begin{equation}
F_{01}  =
\frac{Q_e}{\sqrt{r^4+\frac{Q_2^2}{b^2}}},~~P_{01}=\frac{Q}{r^2}.
\end{equation}
In this work, we consider the magnetically charged case only. The
EBI  black hole solution can be written as
\begin{eqnarray}
f_{BI}(r) &=& 1 - \frac{2M}{r}+ \frac{2 b^2 r^2}{3} \left( 1 -
\sqrt{ 1 + \frac{Q^2}{b^2r^4 }}
 \right)+\frac{ 4 Q^2}{ 3 r}G(r),  \label{fbi}\\
 G'(r)&=&-\frac{1}{\sqrt{ r^4 + \frac{Q^2}{b^2}}}
\end{eqnarray}
where $G'(r)$ denotes the derivative of $G(r)$ with respect to its
argument. Importantly,  comparing (\ref{KSs}) with third term of
(\ref{fbi}) leads to other  correspondence \be \label{appc} m
\leftrightarrow Q^2,~~\omega \leftrightarrow b^2. \ee
 For the EBI
black hole, $G$ takes the form \be
G(r)=\int^\infty_r\frac{ds}{\sqrt{s^4+\frac{Q^2}{b^2}}}=\frac{1}{r}F
\Big[\frac{1}{4},\frac{1}{2},\frac{5}{4};-\frac{Q^2}{b^2r^4}\Big],
\ee where $F$ is the hypergeometric function. In the presence of
negative cosmological constant and  electric charge, its solution
and thermodynamics were given in Refs.\cite{dey,Fernando,mkp}.
 On the other hand,
for the soliton-like solution, it takes the form
 \be
G(r)=-\int^r_0\frac{ds}{\sqrt{s^4+\frac{Q^2}{b^2}}}. \ee For large
$r$, the expansion of $f_{BI}$ is given by \be \label{bis}
f_{BI}\simeq
1-\frac{2M}{r}+\Bigg(\frac{Q^2}{r^2}-\frac{Q^4}{20b^2r^6}+\frac{Q^6}{75b^4r^{10}}-\frac{5Q^8}{832b^6r^{14}}+\cdots\Bigg).
\ee
 At this stage, we note two limiting cases as guided black
holes to study the EBI black hole.  In the limit of  $Q
\rightarrow 0$, this metric function reduces to the Schwarzschild
case (\ref{sch}), while in the limit of $b \rightarrow \infty$ and
$Q \neq 0$, this metric function reduces to the RN black hole
(\ref{RN}). Comparing (\ref{kss}) with (\ref{bis}), one notes that
the correspondence (\ref{appc}) holds roughly.

At infinity, the ADM mass $M_{ADM}$  is obtained from the
condition of $f_{BI}(r_+)=0$~\cite{Breton} \be \label{aas1}
M_{ADM}(r_+,Q,b) = \frac{r_{+}}{2} + \frac{ b^2 r_{+}^3}{3} \left(
1 - \sqrt{ 1 + \frac{Q^2}{ b^2 r_+^4}} \right)+\frac{2Q^2}{3r_+}F
\Big[\frac{1}{4},\frac{1}{2},\frac{5}{4};-\frac{Q^2}{b^2r_+^4}\Big].
\ee This is possible because $M$-term  is a single one  in the
metric function $f_{BI}$.
 On
the other hand, the horizon mass $M_{h}$ is defined to be
 \be \label{aas2}
M_{h}(r_+,Q,b) = \frac{r_{+}}{2} + \frac{ b^2 r_{+}^3}{3} \left( 1
- \sqrt{ 1 + \frac{Q^2}{ b^2 r_+^4}}
\right)-\frac{2Q^2}{3}\int^{r_+}_0
\frac{ds}{\sqrt{s^4+\frac{Q^2}{b^2}}}. \ee In addition, we note
that the soliton mass is obtained as \be \label{EBIon}
M_{sol}=M_{ADM}-M_h=\frac{2Q\sqrt{Qb}K[1/2]}{3},\ee where $K[1/2]$
is the complete elliptical integral of the first kind given by \be
K\Big[\frac{1}{2}\Big]=\frac{\Gamma\Big[\frac{1}{4}\Big]\Gamma\Big[\frac{5}{4}\Big]}
{\Gamma\Big[\frac{1}{2}\Big]}. \ee

Finally, we emphasize  that the horizon mass (\ref{aas2}) is also
derived from the first law of the thermodynamics \be
\label{firstl}dM_h=T_H dS_{BH} \to M_h=\int^{r_+}_0T_H dS_{BH},\ee
where the Hawking temperature is defined by \be
T_{H}(\tilde{r}_+,Q,b)=
\frac{f'_{BI}(\tilde{r}_+)}{4\pi}=\frac{1}{4\pi}
\Bigg[\frac{1}{\tilde{r}_+}+2b^2\tilde{r}_+\Bigg(1-\sqrt{1+\frac{Q^2}{b^2\tilde{r}_+^4}}\Bigg)\Bigg].
\ee In deriving the horizon mass $M_h$, we use the integration
formula \be \int^{r_+}_0
\sqrt{r^4+\frac{Q^2}{b^2}}dr=\frac{r_+^3}{3}\sqrt{1+\frac{Q^2}{b^2r_+^4}}+\frac{2}{3}\frac{Q^2}{b^2}
 \int^{r_+}_0\frac{dr}{\sqrt{r^4+\frac{Q^2}{b^2}}}.\ee
 However, we note that this is possible only for a magnetically
charged EBI black hole, but not for  an electrically charged EBI
black hole~\cite{BGS}. The reason is that if the variation of
electric charge $Q_e$ is taken into account, the first law
(\ref{firstl}) is changed into $dM_h=T_H dS_{BH}+\Phi dQ_e$ at
horizon.

\section{Horizon and ADM masses, and quasilocal energy}
It was shown that the horizon and ADM masses of EBI~\cite{BGS},
colored~\cite{ACS}, and Bardeen  black holes~\cite{BretonS} are
also derived from the first law and the area-law entropy if one
uses magnetic charges. Considering two correspondences
(\ref{naivc}) and (\ref{appc}), we may consider $\omega$ as ``a
pseudo magnetic charge". We remind that  the KS metric function
(\ref{kss}) contains an infinite $m$-dependent terms and thus, one
could not use $f_{KS}=0$ to obtain the horizon and ADM masses as
in the EBI black hole. It suggests that one way to derive these
masses is to use the first law because $\omega$ belongs to a
pseudo magnetic charge.

Now we are in a position to derive the horizon mass for the KS
black hole. Using the Bekenstein-Hawking entropy $S_{BH}=\pi
r_+^2$ and the Hawking temperature in (\ref{oldt}), the horizon
mass is obtained from the first law \be M_{h}(r_+,\omega)=\int T_H
dS_{BH}=\frac{r_+}{2}-\frac{3}{4}\int^{r_+}_0 \frac{dr}{\omega
r^2+1}=\frac{r_+}{2}-\frac{3 \tan^{-1}(\sqrt{\omega}
r_+)}{4\sqrt{\omega}}. \ee On the other hand, the ADM mass is
calculated  to be
 \be M_{ADM}(r_+,\omega)=\frac{r_+}{2}+\frac{3}{4}\int^\infty_{r_+}\frac{dr}{\omega
r^2+1}dr=\frac{r_+}{2}-\frac{3 \tan^{-1}(\sqrt{\omega}
r_+)}{4\sqrt{\omega}}+\frac{3\pi}{8\sqrt{\omega}}. \ee Using the
relation for large $x$ \be
\tan^{-1}x=\frac{x}{1+x^2}F\Big[1,1,\frac{3}{2};\frac{x^2}{1+x^2}\Big]=
\sum^{\infty}_{n=0}\frac{2^{2n}(n!)^2}{(2n+1)!}
\frac{x^{2n+1}}{(1+x^2)^{n+1}},\ee one finds that  the horizon
mass takes a series form \be M_{h}\simeq M_{H}-\frac{\omega
r_+^3}{2(1+\omega r_+^2)^2}\Bigg[1+\frac{4\omega r_+}{5(1+\omega
r_+^2)}+\cdots\Bigg],\ee where the first term represents the Komar
charge\footnote{The Komar charge is originally defined by \be
M_c=\frac{1}{4\pi} \int_{B} {\bf g}\cdot d{\bf s},\ee where ${\bf
g}=-n\nabla (\ln n)$ and $n=\sqrt{-t^at_a}$ with $t^a$ a timelike
Killing vector~\cite{Komar,BD}. In this case, the KS system is a
spatial three-surface $\Sigma$ bounded by a two-surface $B=S^2$.}
at horizon\be
M_{H}=\tilde{m}(r_+)-r_+\tilde{m}'(r_+)=\frac{r_+}{2}-\frac{3r_+}{4(1+\omega
r_+^2)}=2T_H S_{BH},\ee and  remaining terms denote the potential
$V(r_+)$ in the modified Smarr formula.

 Other important quantity
of quasilocal energy is defined by~\cite{BY,LSY} \be
E(r)=\frac{1}{8\pi}\int_{B}d^2x\sqrt{\sigma}(k-k_0),\ee where $B$
is the two dimensional spherical surface $S^2$ with surface area
$A=4\pi r^2$, $k$ is the trace of the extrinsic curvature of $B$,
$\sigma_{ij}$ is the induced metric of $B$, and $k_0$ is the
reference term of the Minkowski spacetimes. Interpreting the Komar
charge and quasilocal energy~\cite{BD}, the former  (gravitational
charge) measures the strength of the gravitational pull exerted by
a body, while the gravitational field energy (quasilocal energy
difference between horizon and infinity) is related to the amount
of curvature of space. For the RN black hole, both quantities are
equals at  horizon. Especially, this  is a nonvariational identity
which relates quantities at horizon and at infinity, in a
different way to the first law of black hole thermodynamics, where
variations  of certain quantities at horizon and at infinity are
related.

For the SSS metric (\ref{sss}), it turns out that  the boundary
condition of $\tilde{m}(r\to \infty)=m$ satisfies asymptotic
flatness. Then, the quasilocal energy inside a spherical surface
of radius $r\ge r_+$ is given by    \be
E(r)=r-r\sqrt{1-\frac{2\tilde{m}(r)}{r}}\ee whose expansion is
given for large $r$ \be E(r)\simeq
m+\frac{m^2}{2r}+\frac{m^3}{2r^2}+\cdots. \ee On the other hand,
the RN metric function provides its quasilocal energy~\cite{Bal}
\be E_{RN}(r)\simeq M+\frac{M^2-Q^2}{2r}+\cdots\ee The quasilocal
energy  provides an interesting difference between the horizon and
infinity  \be
E(r_+)-E(\infty)=\sqrt{m^2(r_+)-\frac{1}{2\omega}}=\frac{r_+}{2}-\frac{1}{4\omega
r_+} \ee where
 \be
 E(\infty)=m.\ee
 For the extremal black hole at
$r_+=r_e=1/\sqrt{2\omega}$, this quantity vanishes. In addition,
the mass parameter at horizon takes the form \be
m(r_+)=\frac{r_+}{2}+\frac{1}{4\omega r_+} \ee which is obviously
different from the horizon mass  $M_h$. Hence,  we identify the
mass parameter $m$ with the quasilocal energy at infinity.

 Furthermore, we show
that the inequality \be E(r_+)-E(\infty)>M_H(r_+)\ee is satisfied
\cite{Bal} and thus, the equality achieves when adding a new term
$\Delta(r_+)$ as \be
E(r_+)-E(\infty)=M_H(r_+)+\Delta(r_+)=2T_H\Big(S_{BH}+\frac{\pi}{\omega}\Big)
\ee with \be \Delta(r_+)=\frac{2\omega r_+^2-1}{4\omega r_+(\omega
r_+^2+1)}=\frac{2\pi}{\omega} T_H.\ee However, the RN black hole
satisfies the equality as \be
E_{RN}(r_+)-E_{RN}(\infty)=\sqrt{M^2-Q^2}=M^{RN}_H(r_+).\ee

\begin{figure}[t!]
   \centering
   \includegraphics{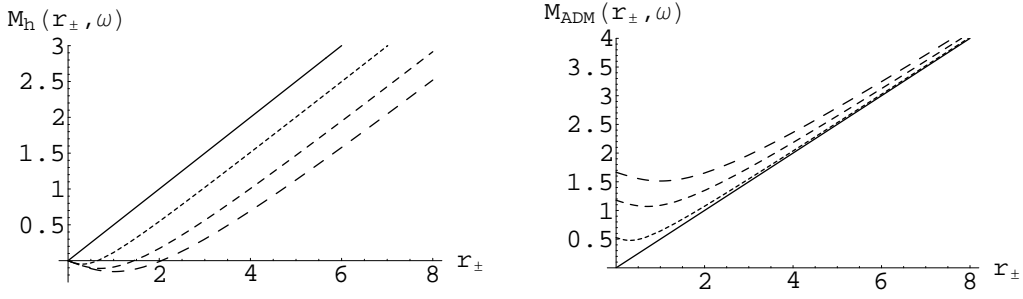}
\caption{Graphs of horizon mass $M_h$ and ADM mass $M_{ADM}$. Left
graph for horizon mass: the upper solid curve represents the
Schwarzschild mass and  three dashed curves denote the KS black
holes with $\omega=0.5,1,5$ from the bottom to  top.
 Right graph for ADM mass: the lower solid curve represents the
Schwarzschild mass and  three dashed curves denote the KS black
holes with $\omega=0.5,1,5$ from the top to bottom } \label{fig.1}
\end{figure}
At this stage, we show that ACS conjecture~\cite{ACS} is satisfied
by the KS black hole. From Fig. 1, the KS horizon mass
$M_h(r_+,\omega)$ is always less than Schwarzschild mass
$M_s=r_+/2$ for any value $\omega$, while the KS ADM mass
$M_{ADM}$ is always greater than the Schwarzschild mass $M_s$. We
note that $M_{h}$ becomes negative for small black holes, while
$M_{ADM}$ is always positive.

Let us calculate  the difference between ADM  and horizon masses
\be \label{KSs1}M_{ADM}-M_h=\frac{3\pi}{8\sqrt{\omega}}, \ee which
may be interpreted as a solitonic mass. Comparing (\ref{EBIon})
with (\ref{KSs1}) leads to a relation \be \omega \sim
\frac{1}{Q^3b}. \ee On the other hand, the positivity of mass
difference may imply a potential instability~\cite{ACS,BGS}. That
is, a perturbation in the initial data will trigger the KS black
hole to decay to a Schwarzschild black hole. Furthermore, the
difference between the KS ADM mass and the Schwarzschild mass
turns out to be positive as \be
M_{ADM}-M_s=\frac{3\pi}{8\sqrt{\omega}}-\frac{3
\tan^{-1}(\sqrt{\omega} r_+)}{4\sqrt{\omega}}>0. \ee On this
basis, one might conjecture that the KS black hole is unstable. In
order to study  a phase transition between two black holes,
however, we use the heat capacity and  free energy.
\section{Phase transitions}
\begin{figure}[t!]
   \centering
   \includegraphics{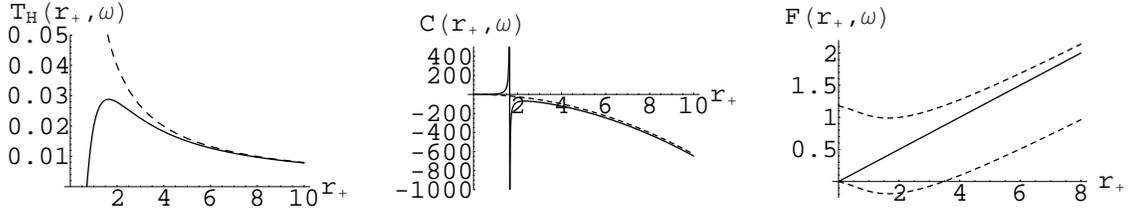}
\caption{ Graphs of temperature, heat capacity, and free energy
with $\omega=1$. Left:  Hawking temperature is zero
($T_H(r_e,\omega)=0$) at the extremal point $r_e=0.71$, while it
is maximum ($T_H=T_m$) at $r_m=1.64$. A dashed curve denotes the
temperature $T_s=\frac{1}{4\pi r_+}$ of the Schwarzschild black
hole. Two are quite different for small black holes. Center:
$C_\omega(r_+,\omega)$ shows a blow-up point
 at $r_m=1.64$, dividing it into $C_\omega>0$ and $C_\omega<0$. Note that $C_\omega(r_e,\omega)=0$.
 A dashed curve denotes
the heat capacity $C_s=-2\pi r_+^2$ of the Schwarzschild black
hole. Right: upper dashed, solid, and lower dashed curves
represent $F_{ADM}(r_+,\omega)$, $F_s=r_+/4$, and
$F_{h}(r_+,\omega)$, respectively.} \label{fig.2}
\end{figure}
The two important  quantities for determining the black hole phase
transition are heat capacity and free energy~\cite{mkp}. The heat
capacity is defined by \be
C_\omega=\Big(\frac{dM_h}{dT_H}\Big)_\omega=\Big(\frac{dM_{ADM}}{dT_H}\Big)_\omega
=-\frac{2\pi r_+^2(\omega r_+^2+1)(2\omega
r_+^2-1)}{2\omega^2r_+^4-5\omega r_+^2-1} \ee which seems to be
different from the old heat capacity in (\ref{oldt}), but its
characteristic is not changed. This quantity is crucial for
testing local thermodynamic stability. As is depicted in Fig. 2,
the heat capacity blows up at the maximum temperature point
$r_+=r_m$, dividing it into $C_\omega>0$ and $C_\omega<0$. The
former case of small black hole is thermodynamically stable, while
the latter of large black hole is thermodynamically unstable, like
the Schwarzschild black hole. This is clear because the KS black
hole has an extremal black hole which is considered to be  a
stable remnant as a final stage of black hole evaporation via
Hawking radiation~\cite{mkpc}. This feature contrasts sharply  to
that of Schwarzschild black hole, showing that a negative heat
capacity makes it hotter and causes the horizon area to shrink.
This process escalates until the horizon collapses rapidly onto
the singularity amid an explosive radiation of quanta.

 The free energy
usually determines a global thermodynamic  stability when
combining with the heat capacity. We have two kinds of free
energies. The free energy at  horizon is defined by \be
F_h=M_h-T_HS_{BH}= \frac{r_+}{2}-\frac{3 \tan^{-1}(\sqrt{\omega}
r_+)}{4\sqrt{\omega}}-\frac{(2\omega r_+^2-1)}{8(\omega r_+^2+1)},
\ee because we consider $\omega$ as a pseudo magnetic charge. On
the other hand, the free energy based on the ADM mass takes a
different form as \be F_{ADM}=M_{ADM}-T_HS_{BH}=
\frac{r_+}{2}-\frac{3 \tan^{-1}(\sqrt{\omega}
r_+)}{4\sqrt{\omega}}+\frac{3\pi}{8\sqrt{\omega}}-\frac{(2\omega
r_+^2-1)}{8(\omega r_+^2+1)}. \ee As is shown in Fig. 2, we find
an important  sequence \be F_{ADM}>F_s>F_h. \ee This implies that
if one uses $F_{ADM}$ instead of $F_h$, one could not explain the
local stability of small black hole, arriving at the extremal
black hole as a stable remnant. In this case, the horizon free
energy $F_h$  is more appropriate for understanding the feature of
KS black hole than the ADM free energy. If one uses the ADM free
energy to describe the phase transition,  the Schwarzschild black
hole is  more stable than  the KS black hole. This implies  that a
perturbation on the KS black hole may induce  the KS black hole to
decay to the  Schwarzschild black hole. This is consistent with
the previous argument based on the mass difference.

\section{Discussions}
Applying the isolated horizon formalism to the KS black hole, we
have found the horizon and ADM mass by using the first law and the
Bekenstein-Hawking entropy. These masses take obviously different
from the mass parameter $m$ in Eq.(\ref{oldt})  obtained from
$f_{KS}=0$. Importantly,  we have identified  the mass parameter
$m$ with  the quasilocal energy at infinity.

This implies that the colored black hole with color index $n$, the
EBI black hole with coupling parameter $b^2$, the Bardeen black
hole with magnetic charge $g$, and the KS black hole with
parameter $\omega$ belong to the same category of  black holes
which  need  a careful study to find the correct thermodynamics.
In this sense, the deformed potential Lagrangian (\ref{depot}) may
be regarded as a non-linear gravity theory of $R$ with coupling
parameter $\omega$.

In order to study  a phase transition between the KS black hole
and Schwarzschild black hole, we introduce the heat capacity and
the free energy. We did not discuss  the black hole phase
transition only by mentioning the mass difference. The heat
capacity shows the local thermodynamic stability, while the free
energy describes the global thermodynamic stability. One basic
difference between two black hole is that the KS black hole has an
extremal black hole with $C_\omega=0$ and $F_h<0$, while the
Schwarzschild black hole has not. In order to explain the stable
remnant at extremal point, we need to use the horizon free energy
but not the ADM free energy.

In conclusion, we have obtained the horizon and ADM masses for the
KS black hole in the deformed HL gravity. In deriving these
masses, we use the first law of thermodynamics and the area-law
entropy. Hence, all thermodynamic quantities are well defined
without any  pathology. A  remaining issue is that the ADM mass is
always greater than the horizon mass, implying that the KS black
hole likely decays to the Schwarzschild black hole. However, this
issue is hard to be accepted because it  unlikely occurs when
considering the extremal black hole as a stable remnant.

\section*{Acknowledgement}
The author thanks Hyung Won Lee and Yong-Wan Kim for helpful
discussions. This work was in part  supported  by Basic Science
Research Program through the National Research  Foundation (NRF)
of Korea funded by the Ministry of Education, Science and
Technology (2009-0086861) and   the NRF grant funded by the Korea
government(MEST) through the Center for Quantum Spacetime (CQUeST)
of Sogang University with grant number 2005-0049409.

\end{document}